\documentclass[aps,prd,dlspace,twocolumn,a4paper]{revtex4}
\usepackage{amsfonts}
\usepackage{epsfig}
\usepackage{graphics}
\usepackage{color}
\newcommand{\beq}{\begin{equation}}
\newcommand{\eeq}{\end{equation}}
\newcommand{\bqa}{\begin{eqnarray}}
\newcommand{\eqa}{\end{eqnarray}}

\newcommand{\chic}{\chi_{c}}

\newcommand{\eps}{\epsilon_s}

\newcommand{\jpsi}{J / \psi}

\newcommand{\pt}{p_{{}_T}}

\newcommand{\spt}{S(p_{{}_T})}
\newcommand{\sinpt}{ \langle S(p_{{}_T})\rangle}
\newcommand{\sincl}{\langle S^{{}^{\mbox{incl}}} \rangle}
\newcommand{\sdir}{\langle S^{{}^{\mbox{dir}}} \rangle}

\begin{document}
\title{Equation of states and charmonium suppression in Heavy ion collisions}
 \author{Indrani Nilima $^{a}$}
\email{nilima.ism@gmail.com}
\author{Vineet Kumar Agotiya$^{a}$}
\email{agotiya81@gmail.com}
 \affiliation{$^a$Department of Physics, Central University of Jharkhand
 Ranchi, India, 835 205}

\begin{abstract}
The present article is the follow-up of our work Bottomonium suppression in quasi-particle model, where we have extended the study
for charmonium states using quasi-particle model in terms of quasi-gluons and quasi quarks/antiquarks as a equation of state.
By employing medium modification to a heavy quark potential thermodynamic observables
{\em viz.} pressure, energy density, speed of sound etc. have been calculated which 
nicely fit with the lattice equation of state for gluon, massless and as well {\em massive} flavored plasma.
For obtaining the thermodynamic observables we employed the debye mass in the quasi particle picture. We extended the quasi-particle model
to calculate charmonium suppression in an expanding, dissipative
strongly interacting QGP medium (SIQGP). We obtained the suppression pattern for charmonium states
with respect to the number of participants at mid-rapidity  and compared it with the experimental data
(CMS JHEP) and (CMS PAS) at LHC energy (Pb+Pb collisions, $\sqrt{s_{NN}}$= $2.76$ TeV).

\end{abstract}
\maketitle


\section{Introduction}
The primary goal of heavy-ion experiment at the RHIC and the LHC is to search a new state of matter, i.e. the Quark Gluon Plasma. To study the properties of the Quark 
Gluon Plasma (QGP)heavy quarks are considered to be a suitable tool. Initially, the heavy quarks can be calculated in pQCD, which are produced
in primary hard N N collisions~\cite{nason}.The charmonia is a bound states of charm ($c$) and anti-charm ($\overline c$), which is an extremely broad
and interesting field of investigation~\cite{Bra11}. Charmonium states can have smaller sizes than hadrons (down to a few tenths of a fm) and 
large binding energies ($> 500$ MeV)~\cite{Eic80}. In ultrarelativistic heavy-ion collisions, it has been realized that early ideas associating 
with charmonium suppression with the deconfinement transition~\cite{tmatsui} are less direct than originally hoped for~\cite{zhao,blaschk,kulberg,stachel}.

   At sufficiently large energy densities,lattice QCD calculations predict that hadronic matter undergoes a phase transition of deconfined quarks and gluons, called Quark
Gluon Plasma (QGP). In order to reveal the existence and to analyze the properties of this phase transition several research in this direction has been done. In the
high-energy heavy-ion collision field, the study of charmonium production and supression is the most interesting investigations, since, the charmonium yield would be
suppressed in the presence of a QGP due to color Debye screening~\cite{tmatsui}. 

In heavy-ion collisions, charmonium suppression study have been carried out first at the the Super Proton Synchrotron (SPS) by the NA38~\cite{NA38,NA50,NA51},
and NA60~\cite{NA60} then at the Relativistic Heavy Ion Collider (RHIC) by the PHENIX experiment at $\sqrt{s_{NN}}$= 200 GeV~\cite{adare}. The suppression is defined by 
the ratio of the yield measured in heavy-ion collisions
and a reference, called the nuclear modification factor $R_{AA}${~\cite{expt1}  and it is considered as a suitable probe to identify
the nature of the matter created in heavy ion collisions. At high temperature, Quantum chromodynamics (QCD) is believed to be in quark gluon plasma (QGP) phase, 
which is not an ideal gas of quarks  and gluons, but rather a liquid having very low shear viscosity to entropy density ($\eta/s$) ratio~\cite{star,vis1,shur,son}.   

This strongly suggest that QGP may lie in the non-perturbative domain of QCD which is very hard to address both analytically and computationally.
Similar conclusion about QGP and perfect fludity of QGP have been reached from recent lattice studies and from the AdS/CFT studies~\cite{son}, spectral functions and 
transport coefficients in lattice QCD~\cite{satz} and studies based on classical strongly coupled plasmas~\cite{shur1,shur2}, which predict that the 
equation of state (EoS) is interacting even at $T\sim 4 T_c$~\cite{leos,cheng,karsch,gavai}.

The bag model, confinement models, quasi-particle models, are the several models for studying the EoS of strongly interacting quark gluon plasma\cite{sqgp,banscqgp} etc.  
Here in our analysis we are using  quasi-particle debye mass~\cite{pe.1} where equation of state was derived with temperature dependent parton masses 
and bag constant\cite{lh.1,pe.2}, with effective degrees of freedom~\cite{s.1}, etc. All of them claim to explain lattice results, either by adjusting free 
parameters in the model or by taking lattice data on one of the thermodynamic quantity as an input and predicting other quantities. However, physical 
picture of quasi-particle model and the origin of various temperature dependent quantities are not clear yet~\cite{rh.2}.   
In strongly interacting QGP~\cite{sqgp1,sqgp2,sqgp3}, one considers all possible hadrons even at $T\,>\,T_c$ and try to explain non-ideal behavior of QGP near $T_c$. 
Recently, an equation of state for strongly-coupled plasma has been inferred by utilizing the understanding from strongly coupled QED plasma~\cite{ba_cor.1} which 
fits lattice data well. It is implicitly assumed that, once the charmonium dissociates,the heavy quarks hadronize by combining with light quarks only~\cite{alberico}.
About $60\%$ of the observed $J/\psi$'s are directly produced in a hadronic collisions ,the remaining stemming from the decays of $\chi_c$ and the $\psi'$ , excited
charmonium states . Since each $c\bar{c}$ bound state dissociates at a different temperature, a model of {\em sequential suppression} was developed, with the aim of 
reproducing the charmonium suppression pattern in the heavy ion collision \cite{satz2,diga1,diga2,kar,Kar97}. A suppressed yield of quarkonium in the dilepton spectrum,
measured in experiments~\cite{jpsi_sps,exp} was proposed as a signature of QGP formation. To determine quarkonium spectral functions at finite temperature there are 
mainly two theoretical lines of studies are potential models~\cite{pnrqcd,wong} and lattice QCD~\cite{satz,lattice}. 
 
The central theme of our work is that the potential which we are considering in the deconfined phase could have a nonvanishing confining (string) term, in addition
to the Coulomb term~\cite{prc-vineet} unlike Coulomb interaction alone in the aforesaid model~\cite{banscqgp}. By incorporating this potential we had calculated the
thermodynamic variables {\em viz} pressure, energy density, speed of sound etc. Our results match nicely with the lattice results of gluon \cite{leos}, 
2-flavor (massless) as well as 3-flavor (massless) QGP \cite{ka.2}. There is also an agreement with (2+1) (two massless and one is massive) and 4 flavoured lattice 
results too. Motivated by the agreement with lattice results, we employ our equation of state (using quasi-particle Debye mass) to study the Charmonium suppression 
in an expanding plasma in the presence of viscous forces. Here in this work we are not considering the bulk viscosity. This issue will be taken in consideration in near 
future. The $R_{AA}$ of prompt and nonprompt $J/\psi$ has been measured separately by CMS in 
bins of transverse momentum, rapidity and collision centrality~\cite{expt1}. We have compared our results with the experimental data
(CMS JHEP)~\cite{expt1} and (CMS PAS)~\cite{expt2} in Pb+Pb collision at LHC energy and found $\sincl$ is closer to the the experimental results.

In our previous work~\cite{aihep_nilima}, we had calculated the plasma parameter, pressure, energy density and 
speed of sound for only 3-flavor QGP and finally studied the sequential suppression for bottomonium states at the LHC energy in a longitudinally 
expanding partonic system for only $\eta/s=0.08$ because the experimental data 
is available  only  for ADS/CFT case. In this present article we have extended our previous work for charmonium states for all 3-flavors by
using quasi-particle model in terms of quasi-gluons and quasi quarks/antiquarks as a equation of state. Here, we had considered three values of
the shear viscosity-to-entropy density ratio to see the effects of nonzero values of the shear viscosity on the expansion. The first one
is from perturbative QCD calculations where $\eta/s$ is =0.3 near $T\sim T_c \sim 2T_c$. The second one is from AdS/CFT studies where 
$(\eta/s)= 1/4\pi\sim 0.08$. Finally we consider $\eta/s$=0 (for the ideal fluid) for the sake of comparison. These three ratios has been used only 
for the charmonium states for both EoS1 and EoS2.

The paper is organized as follows. In Sec.II., we briefly discuss our recent work on medium modified potential in isotropic medium. In the subsection
II (A) we study the Effective fugacity quasi-particle model(EQPM).
In section III we studied about binding energy and dissociation temperature of $J/\psi$, $\psi^\prime$ and $\chi_c$ state considering isotropic medium.
 Using this effective potential and by incorporating quasi-particle debye mass, we have then developed the equation of state for strongly interacting matter and have
 shown our results on pressure,energy density and speed of sound etc. along with the lattice data in Sec.IV. In Sec.V, we have employed the aforesaid equation of state
to study the suppression of charmonium in the presence of viscous forces and estimate the survival probability in a longitudinally expanding QGP. 
Results and discussion will be presented in Sec.VI and finally, we conclude in Sec.VII.

\section{Medium modified effective potential and fugacity quasi-particle model}
The interaction potential between a heavy quark and antiquark gets modified in the presence of a medium. The static interquark potential plays vital
role in understanding the fate of quark-antiquark bound states in the hot QCD/QGP medium.
In the present analysis, we preferred to work with the Cornell potential~\cite{Eichten:1978tg,Eichten:1979ms}, that 
contains the Coulombic as well as the string part given as,
\begin{equation}
{\text V(r)} = -\frac{\alpha}{r}+\sigma r,
\label{eq:cor} 
\end{equation}
Here, $r$ is the effective radius of the corresponding quarkonia state,  $\alpha$ is  the strong coupling constant
and $\sigma$ is the string tension. The in-medium modification can be obtained in the Fourier space by dividing the heavy-quark 
potential from the medium dielectric permittivity, $\epsilon({k})$ as,
 \begin{equation}
 \grave{V}(k)=\frac{{\bar{\text V}}(k)}{\epsilon(k)}.
 \end{equation}
 where ${\bar{\text V}}(k)$, is the Fourier transform of ${\text V(r)}$, shown in Eq.~\ref{eq:cor}, given as,
\begin{equation}
{\bar{\text V}}(k)= -\sqrt\frac{2}{\pi}\bigg(\frac{\alpha}{k^2}+2\frac{ \sigma}{k^4}\bigg).
\label{eqn4}
\end{equation}
and $\epsilon(k)$ is the dielectric permittivity which is obtained from the static limit of the longitudinal part of gluon 
self-energy\cite{Schneider:prd66}
\begin{eqnarray}
\label{eqn5}
\epsilon(k)=\left(1+\frac{ \Pi_L (0,k,T)}{k^2}\right)\equiv
\left( 1+ \frac{m_D^2}{k^2} \right).
\label{eqn5}
\end{eqnarray}

Next, substituting Eq.(\ref{eqn4}) and Eq.(\ref{eqn5}) into Eq.(\ref{eq:cor}) and evaluating the inverse FT, we obtain r-dependence of the
 medium modified potential~\cite{ldevi}:
\begin{eqnarray}
\label{eq6}
{\bf V}(r,T)&=&\left( \frac{2\sigma}{m^2_D}-\alpha
\right)\frac{\exp{(-m_Dr)}}{r}\nonumber\\
&-&\frac{2\sigma}{m^2_Dr}+\frac{2\sigma}{m_D}-\alpha m_D
\end{eqnarray}

In the limiting case  $r>>1/m_D$, the dominant terms in the potential are
 the long range Coulombic tail and  $\alpha m_D$. The potential will look as,
\begin{eqnarray}
\label{lrp}
{V(r,T)}\sim -\frac{2\sigma}{m^2_Dr}-\alpha m_D
\end{eqnarray}. 

Now we employ the Debye mass computed from the effective fugacity quasi-particle model (EQPM)~\cite{chandra1,chandra2} to determine
the dissociation temperatures for the charmonium states in isotropic medium computed for EoS1 and EoS2 respectively and develop the equation of state for strongly
interacting matter. 
\begin{figure*}
\vspace{-2mm}
\includegraphics[scale=.35]{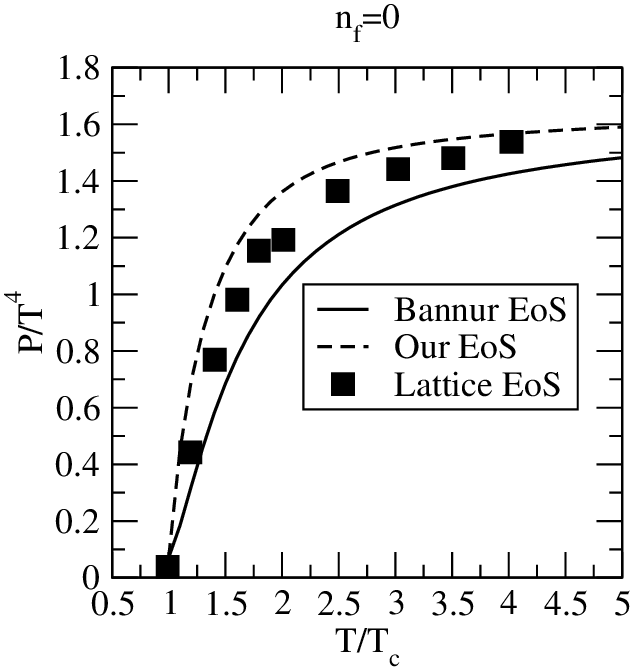}
\hspace{4mm}
\includegraphics[scale=.35]{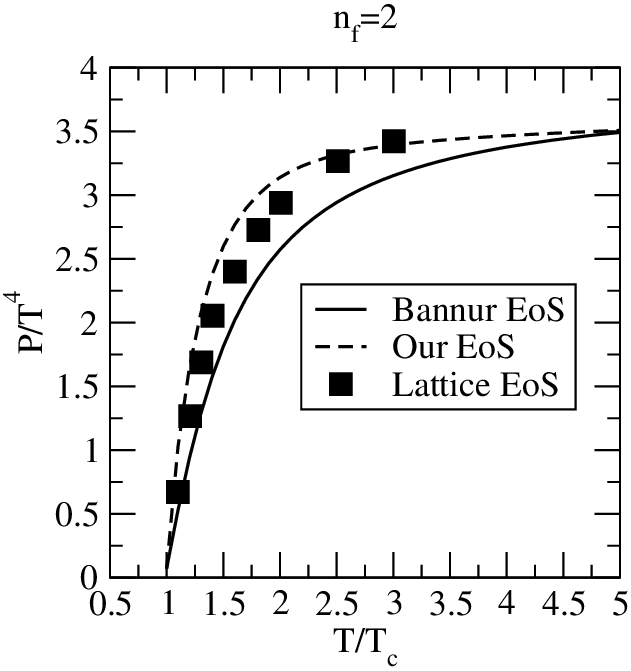}
\hspace{4mm}
\includegraphics[scale=.35]{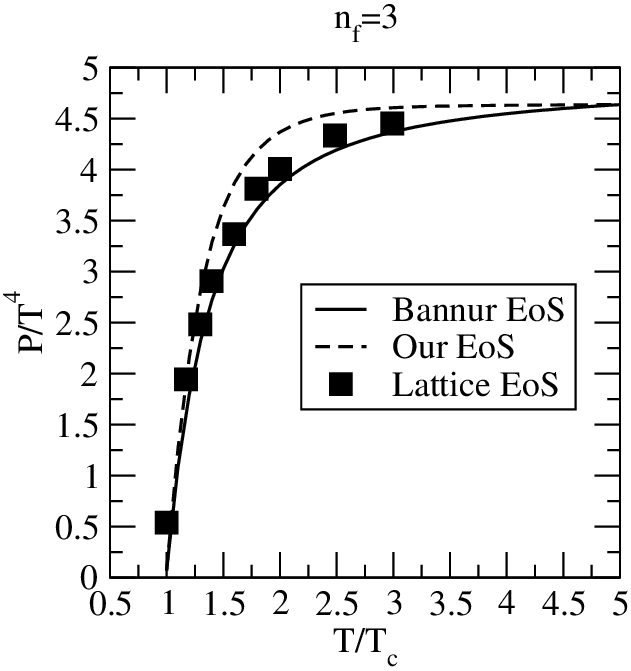}
\vspace{-1mm}
\caption{Plots of $ P/T^4 $ as a function of $T/T_c$ for Bannur EoS,
Our EoS (using quasi-particle Debye mass), and lattice results~\cite{banscqgp, boyd} for pure gauge 
(extreme left figure), 2-flavor QGP (middle figure) and 3-flavor QGP (extreme 
right figure) for EOS1~\cite{zhai,arnold}. In each figure, solid line represents the results obtained 
from Bannur EoS, dashed line represents the results from Our EoS and 
diamond symbols represent lattice results.}
\vspace{30mm}
\end{figure*}
\begin{figure*}
\vspace{-2mm}
\includegraphics[scale=.35]{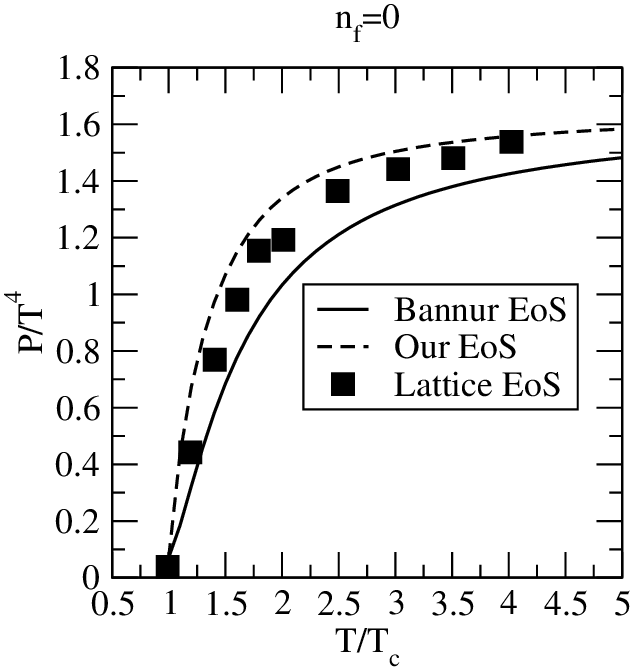}
\hspace{4mm}
\includegraphics[scale=.35]{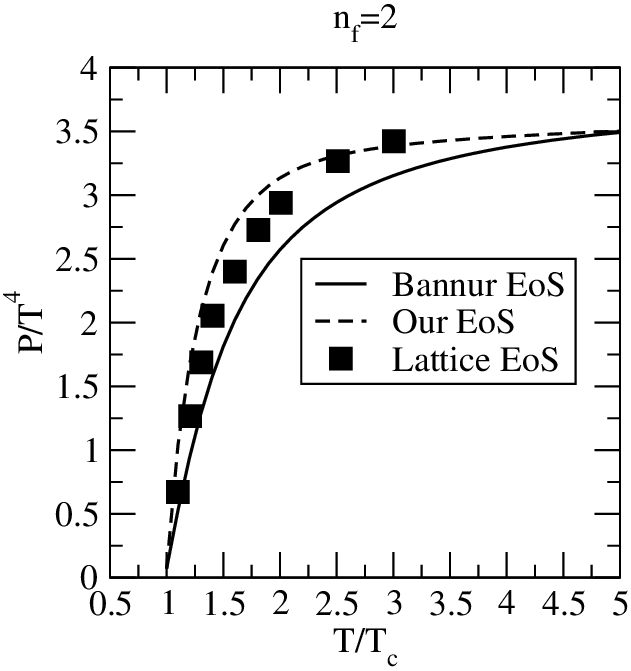}
\hspace{4mm}
\includegraphics[scale=.35]{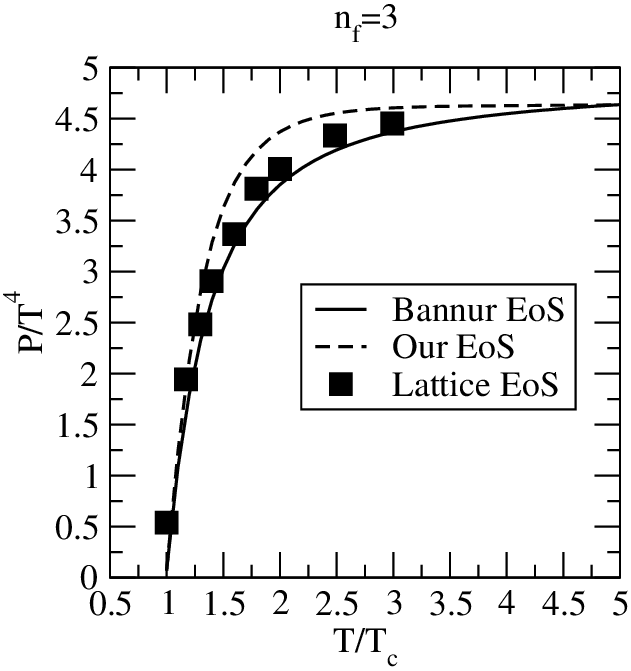}
\vspace{-1mm}
\caption{Plots of $ P/T^4 $ as a function of $T/T_c$ for Bannur EoS, 
Our EoS (using quasi-particle Debye mass) and lattice results~\cite{banscqgp, boyd} for pure gauge 
(extreme left figure), 2-flavor QGP (middle figure) and 3-flavor QGP (extreme 
right figure) for EOS2~\cite{kajantie}. The notations are same as Figure1.}
\vspace{26mm}
\end{figure*}
\begin{figure*}
\vspace{-1mm}
\includegraphics[scale=.40]{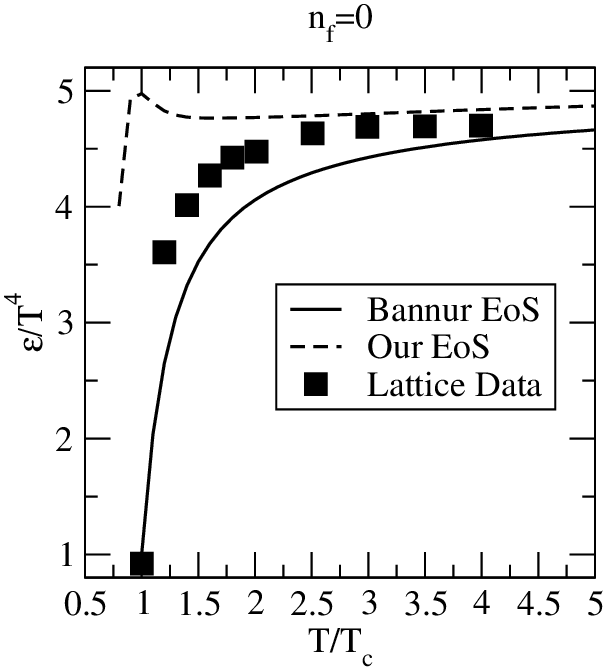}
\hspace{4mm}
\includegraphics[scale=.40]{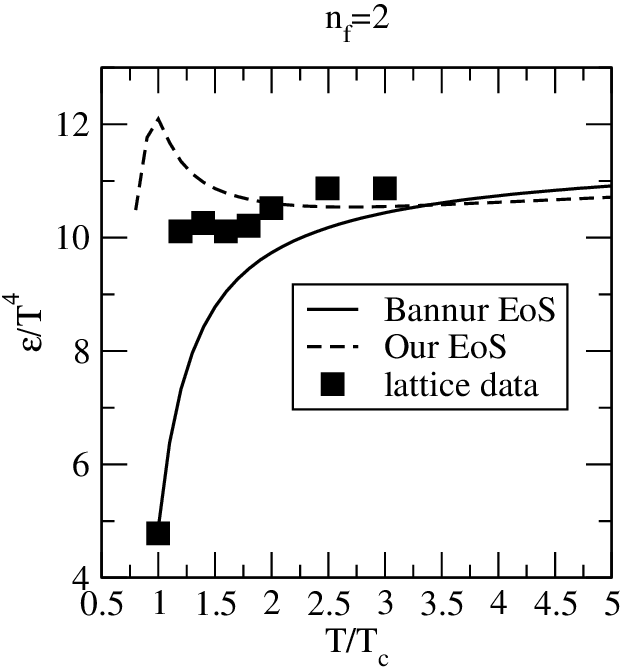}
\hspace{4mm}
\includegraphics[scale=.40]{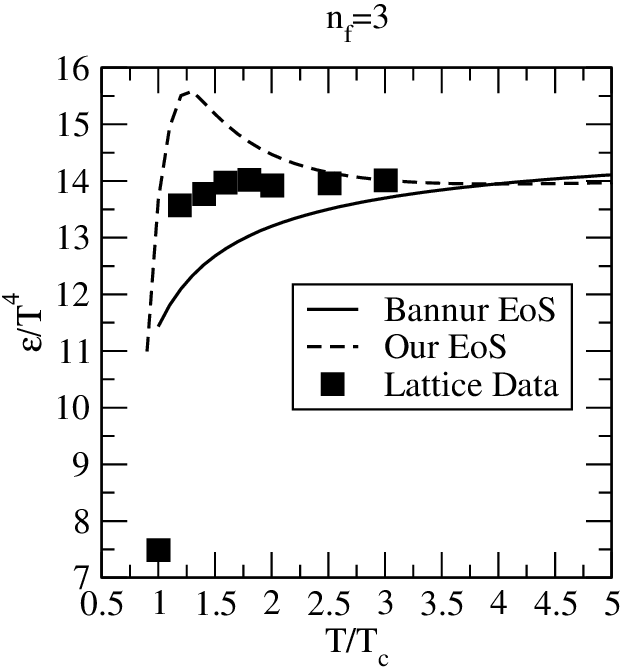}
\vspace{-1mm}
\caption{Plots of $\varepsilon/ T^4 $ as a function of $T/T_c$ for 
Our EoS (using quasi-particle Debye mass) and lattice results~\cite{banscqgp, boyd} for pure gauge 
(extreme left figure), 2-flavor QGP (middle figure) and 3-flavor QGP (extreme 
right figure) for EoS1~\cite{zhai,arnold}. The notations are same as Figure1.}
\vspace{32mm}
\end{figure*}
\begin{figure*}
\vspace{-1mm}
\includegraphics[scale=.40]{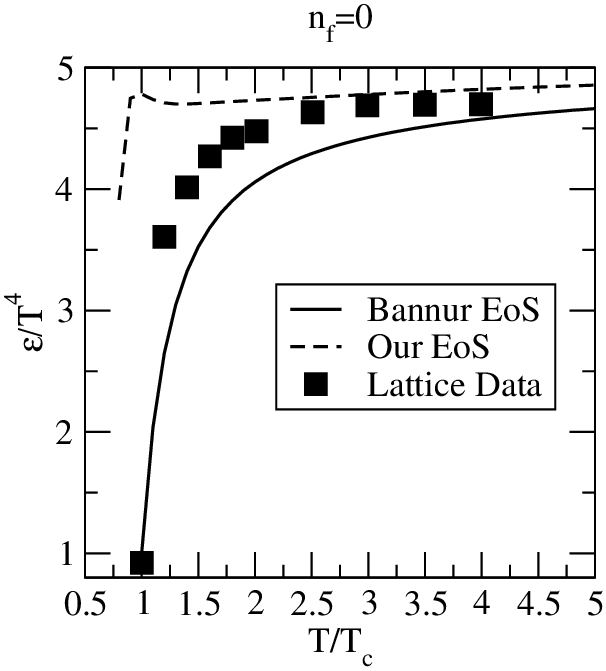}
\hspace{4mm}
\includegraphics[scale=.40]{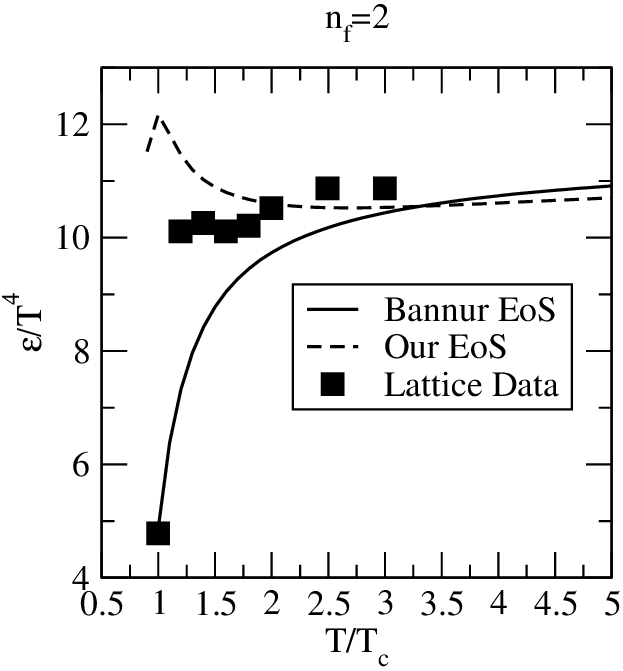}
\hspace{4mm}
\includegraphics[scale=.40]{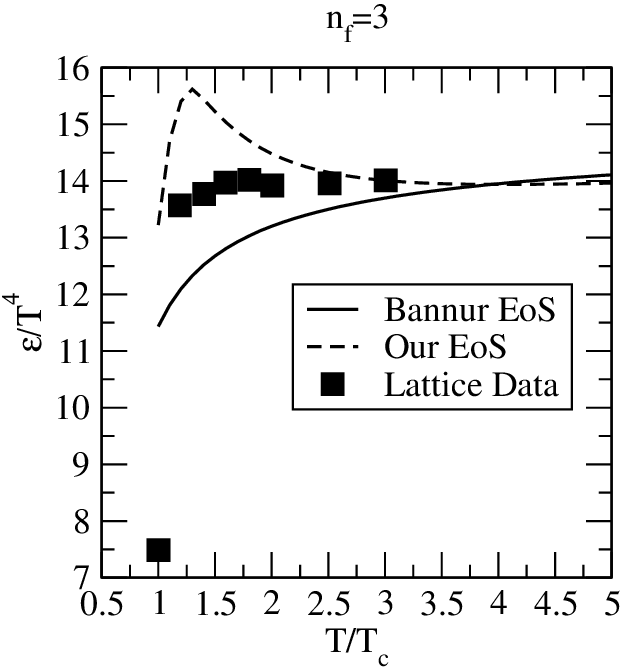}
\vspace{-1mm}
\caption{Plots of $\varepsilon/ T^4 $ as a function of $T/T_c$ for 
Our EoS (using quasi-particle Debye mass) and lattice results~\cite{banscqgp, boyd} for pure gauge 
(extreme left figure), 2-flavor QGP (middle figure) and 3-flavor QGP (extreme 
right figure) for EoS2~\cite{kajantie}. The notations are the same as in Figure1.}
\vspace{26mm}
\end{figure*}
The Debye mass, $m_D$ is defined in terms of the equilibrium (isotropic) distribution function as, 
\begin {equation}
\label{debye}
 m_D^2 \equiv -g^2 
 \int  \frac{{\rm d}^3\bar{\vec{p}}}{(2\pi)^3} \, 
 \frac{{\rm d}f_{eq}(\bar{p})}{{\rm d}\bar{p}}.
\end {equation}
where, $f_{eq}$ is taken to be a combination of ideal Bose-Einstein
 and Fermi-Dirac distribution functions as~\cite{rebhan}, and is given by:
\begin {equation}
\label{eq8}
f_{eq}=2 N_c f_{g}(\vec{p}) + 2 N_f (f_q(\vec{p}) + f_{\bar{q}}(\vec{p})).
\end {equation}
Here, $f_g$ and $f_q$ are the quasi-parton thermal distributions, $N_c$ denotes the number of colors and $N_f$ the number of flavors.

Now, we obtain quasi particle debye mass for full QCD/QGP medium by considering quasi parton distributions  and EoS1 is the $O(g^5)$ hot
 QCD \cite{zhai,arnold} and EoS2 is the $O(g^6\ln(1/g)$ hot QCD EoS \cite{kajantie} in the quasi-particle description \cite{chandra1,chandra2} respectively. 
 
\section{Binding energy and Dissociation Temperature}
Binding energy is defined as the distance between peak position and continuum threshold at finite temperature. The medium modified  potential have the similar appearance to the hydrogen atom problem~\cite{matsui}.Therefore to get the binding energies with medium modified potential we need to solve the Shr\"{o}dinger equation numerically.
 The solution of the Schr\"{o}dinger equation gives the eigenvalues for the 
ground states and the first excited states in charmonium
 ($\jpsi$, $\psi^\prime$ etc.) and bottomonium
 ($\Upsilon$, $\Upsilon^\prime$ etc.)spectra :
\begin{eqnarray}
\label{bind1}
E_n=-\frac{1}{n^2} \frac{m_Q\sigma^2}{m^4_D},
\end{eqnarray}
where $m_Q$ is the mass of the heavy quark. 

In our analysis,the quark masses $m_Q$, as $m_{J/\psi} = 3.09~$GeV,
$m_{\psi^\prime} = 3.68~$GeV  and $m_{\chi_c} = 3.73$GeV, as calculated in ~\cite{Aulchenko:2003qq} and the string tension ($\sigma$) is taken as $0.184GeV^2$. 

We are listed the values of dissociation temperature in Table I and II for the charmonium states $J/\psi$, $\psi^\prime$ and $\chi_c$ for EoS1 and EoS2 respectively, and also seen that $\psi^\prime$ dissociates at lower temperatures as compared to $J/\psi$ and $\chi_c$ for both the EoS.
\begin{figure*}
\vspace{-1mm}
\includegraphics[scale=.40]{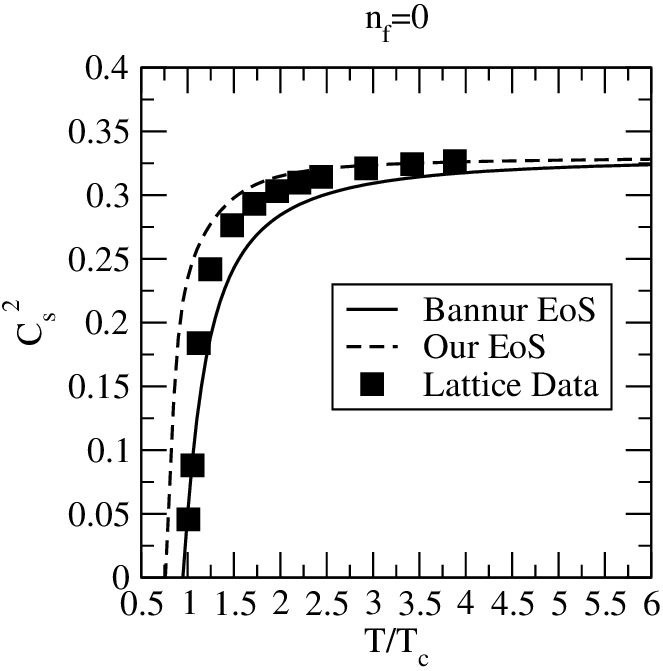}
\hspace{4mm}
\includegraphics[scale=.40]{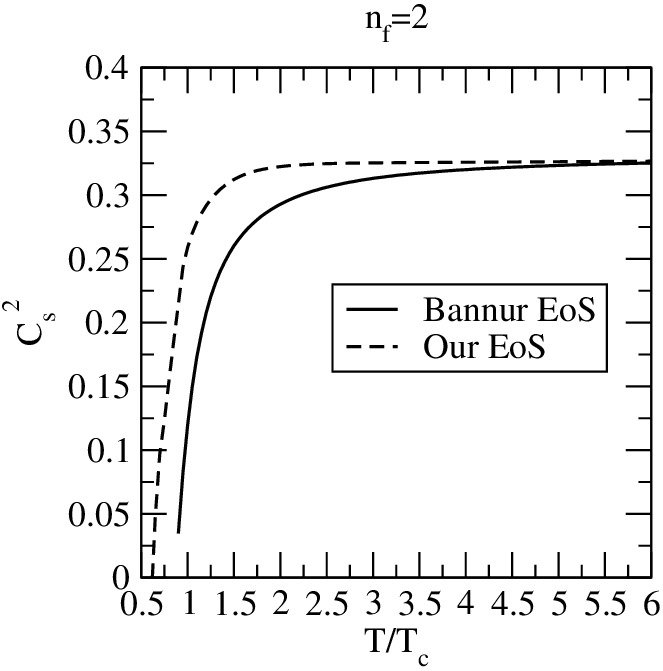}
\hspace{4mm}
\includegraphics[scale=.40]{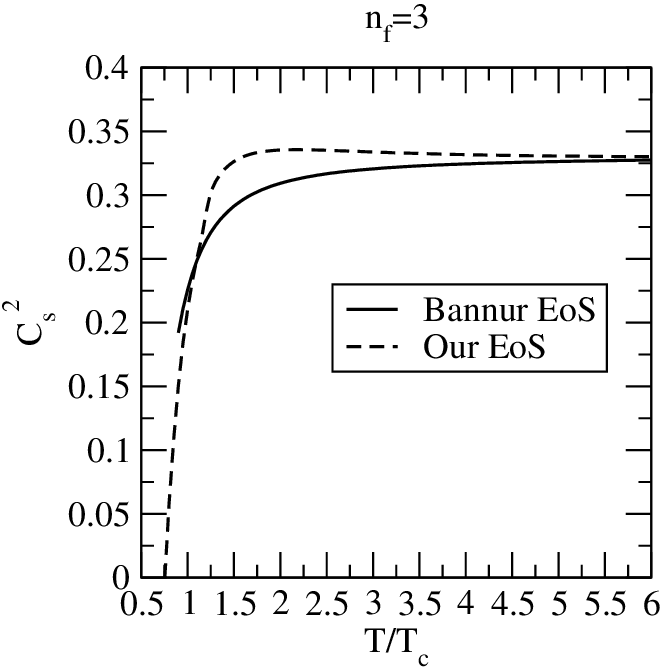}
\vspace{-1mm}
\caption{Plots of $c_s^2$ as a function of $T/T_c$ for Bannur EoS, 
Our EoS (using quasi-particle Debye mass) for pure gauge 
(extreme left figure), 2-flavor QGP (middle figure) and 3-flavor QGP (extreme 
right figure) for EOS1~\cite{zhai,arnold}.} 
\vspace{32mm}
\end{figure*}
\begin{figure*}
\vspace{-1mm}
\includegraphics[scale=.40]{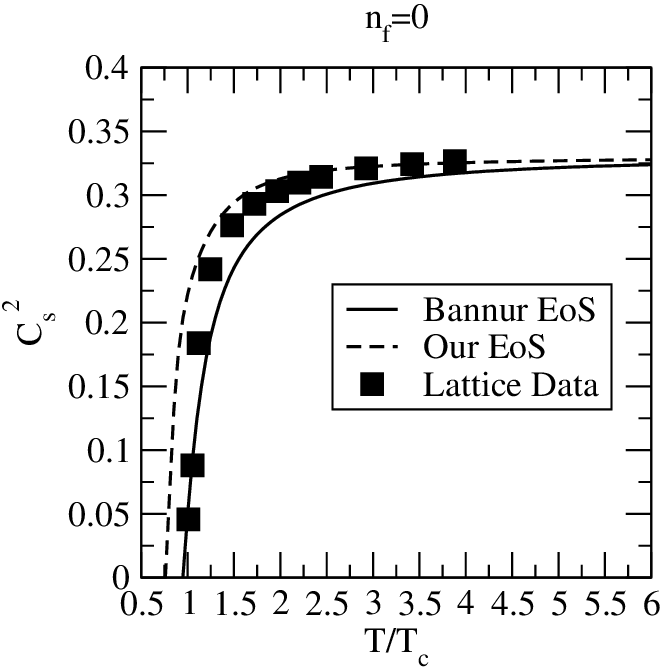}
\hspace{4mm}
\includegraphics[scale=.40]{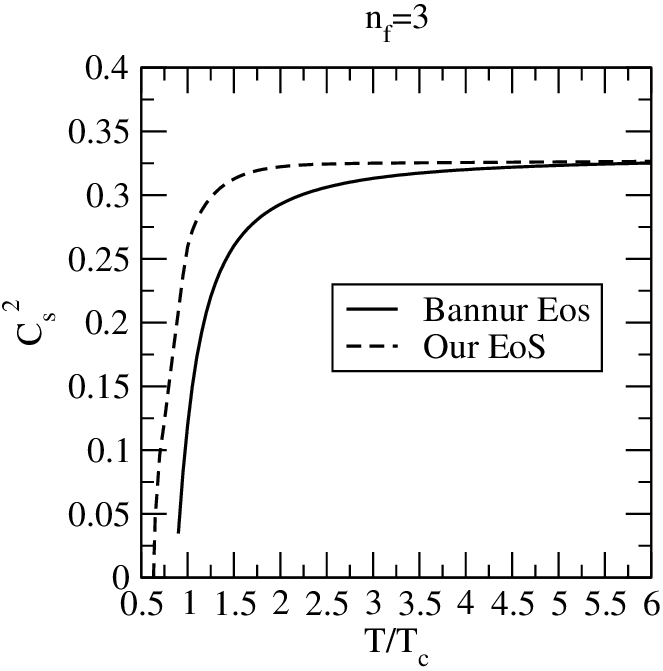}
\hspace{4mm}
\includegraphics[scale=.40]{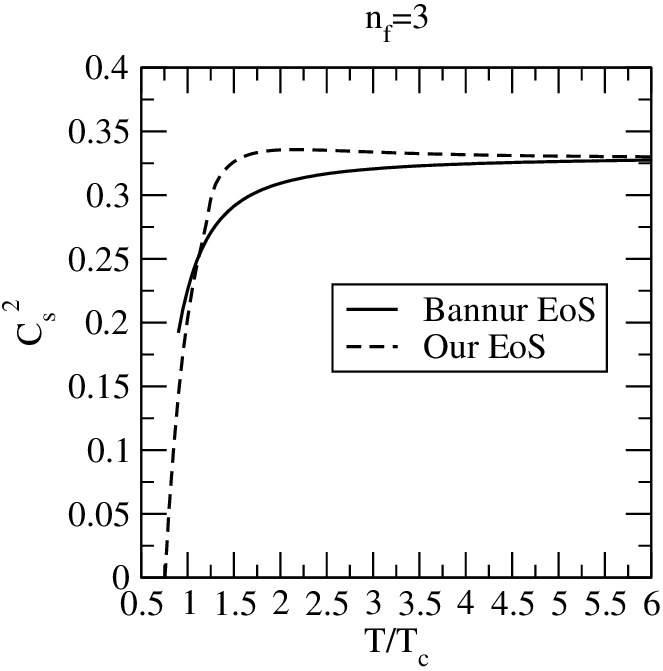}
\vspace{-1mm}
\caption{Plots of $c_s^2$ as a function of $T/T_c$ for 
Bannur EoS, Our EoS (using quasi-particle Debye mass) for pure gauge 
(extreme left figure), 2-flavor QGP (middle figure) and 3-flavor QGP (extreme 
right figure) for EOS2~\cite{kajantie}.} 
\vspace{26mm}
\end{figure*}
\begin{figure*}
\vspace{-1mm}
\includegraphics[scale=.35]{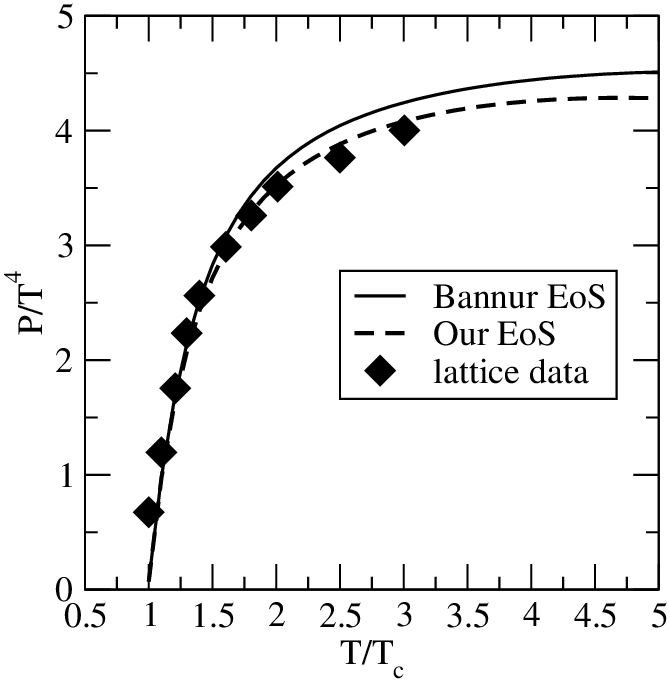}
\hspace{4mm}
\includegraphics[scale=.35]{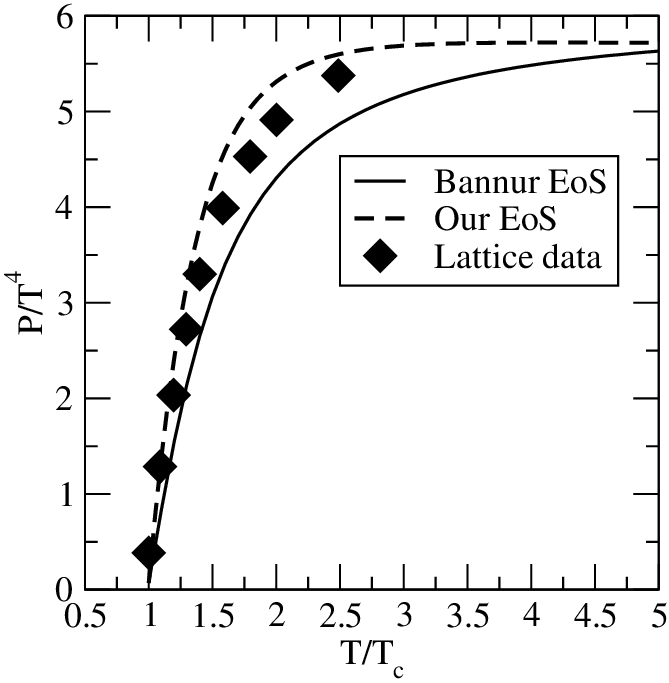}
\hspace{4mm}
\includegraphics[scale=.35]{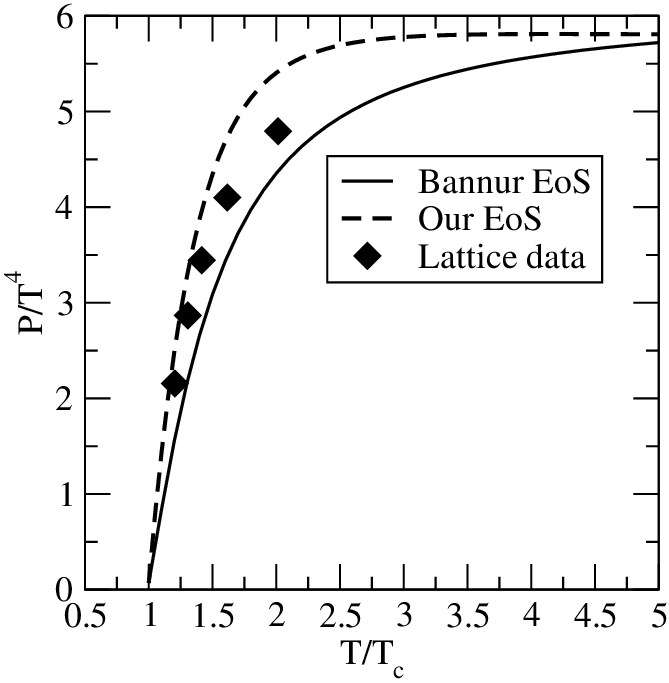}
\vspace{-1mm}
\caption{variation of $P/T^4 $ as a function of $T/T_c$ for Bannur,
Our EoS (using quasi-particle Debye mass) and lattice results~\cite{banscqgp, boyd} for two massless and one massive (2+1) extremely left, middle and extremely right 
figure for 4-flavour QGP for 
two different masses, $m/T$=0.4 and 0.2, respectively for EoS1~\cite{zhai,arnold}. The notations are the same as in Fig.1.}
\vspace{36mm}
\end{figure*}
\begin{figure*}
\vspace{-1mm}
\includegraphics[scale=.35]{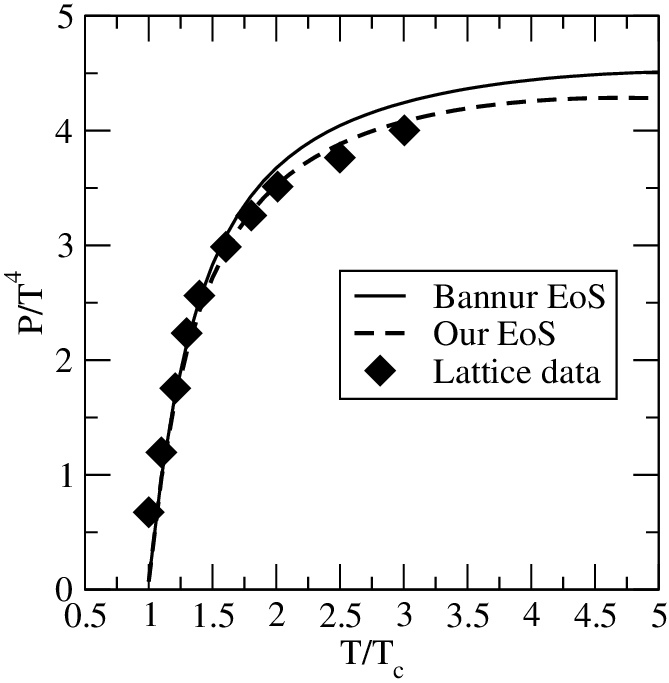}
\hspace{4mm}
\includegraphics[scale=.35]{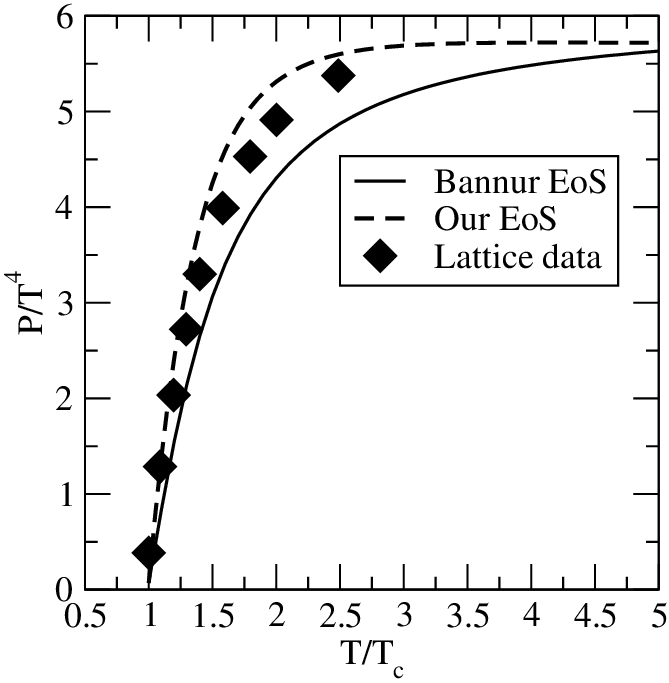}
\hspace{4mm}
\includegraphics[scale=.35]{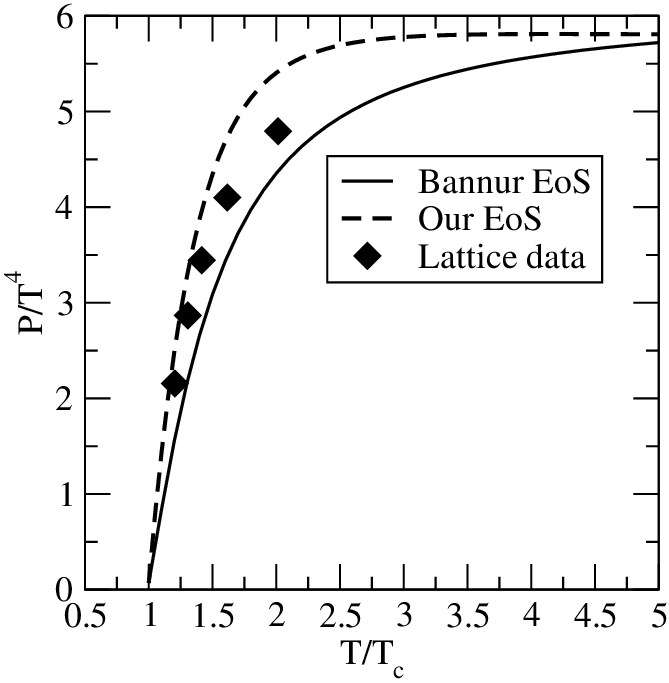}
\vspace{-1mm}
\caption{variation of $P/T^4 $ as a function of $T/T_c$  for Bannur, 
Our EoS (using quasi-particle Debye mass) and lattice results~\cite{banscqgp, boyd} for two
massless and one massive (2+1) extremely left, middle and extremely right for 4-flavour QGP for 
two different masses, $m/T$=0.4 and 0.2, respectively for EoS2~\cite{kajantie}. The notations 
are the same as in Fig.1.}
\vspace{26mm}
\end{figure*}
\begin{figure*}
\vspace{-1mm}
\includegraphics[scale=.39]{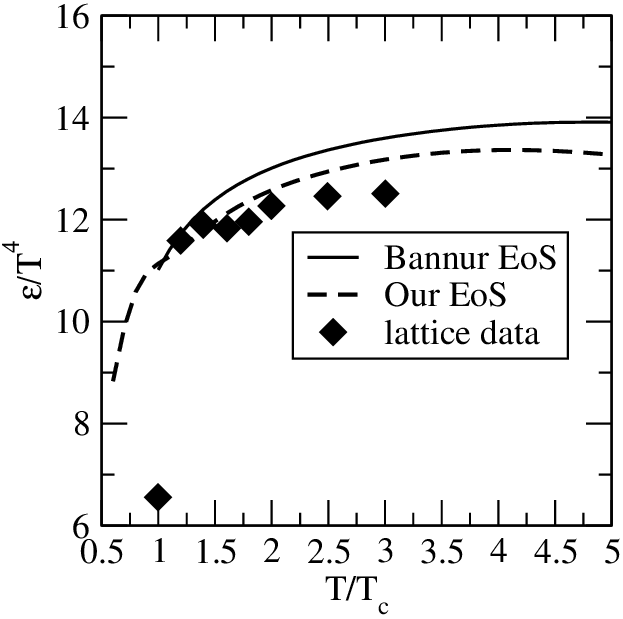}
\hspace{4mm}
\includegraphics[scale=.35]{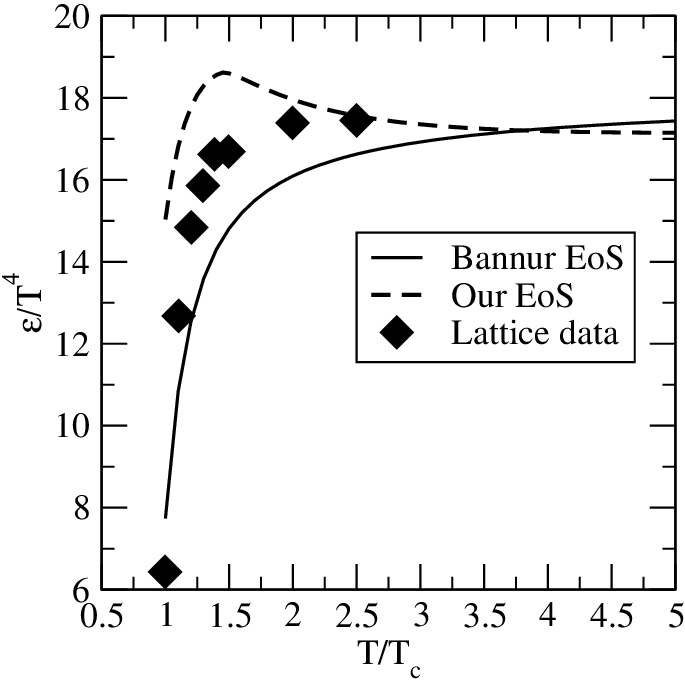}
\hspace{4mm}
\includegraphics[scale=.35]{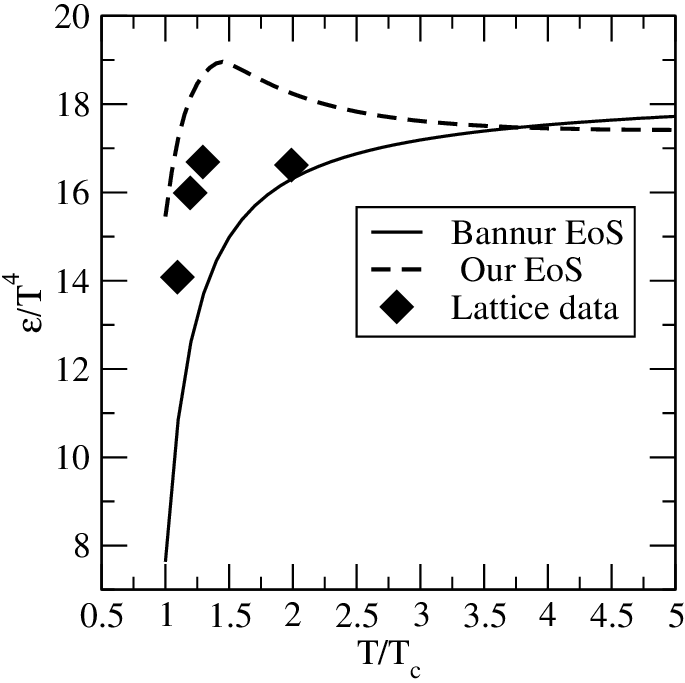}
\vspace{-1mm}
\caption{Variation of $\varepsilon/ T^4 $ as a function of $T/T_c$ for Bannur, 
Our EoS (using quasi-particle Debye mass) and lattice results~\cite{banscqgp, boyd} for two
massless and one massive (2+1) extremely left, middle and extremely right for 4-flavour QGP for 
two different masses, $m/T$=0.4 and 0.2, respectively for the EoS1~\cite{zhai,arnold} 
where the notations are the same as in Fig.1.}
\vspace{48mm}
\end{figure*}
\begin{figure*}
\vspace{1mm}
\includegraphics[scale=.39]{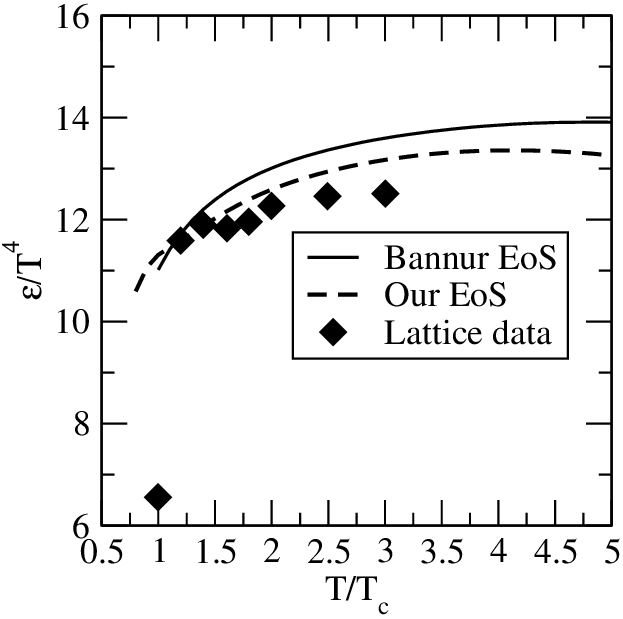}
\hspace{4mm}
\includegraphics[scale=.35]{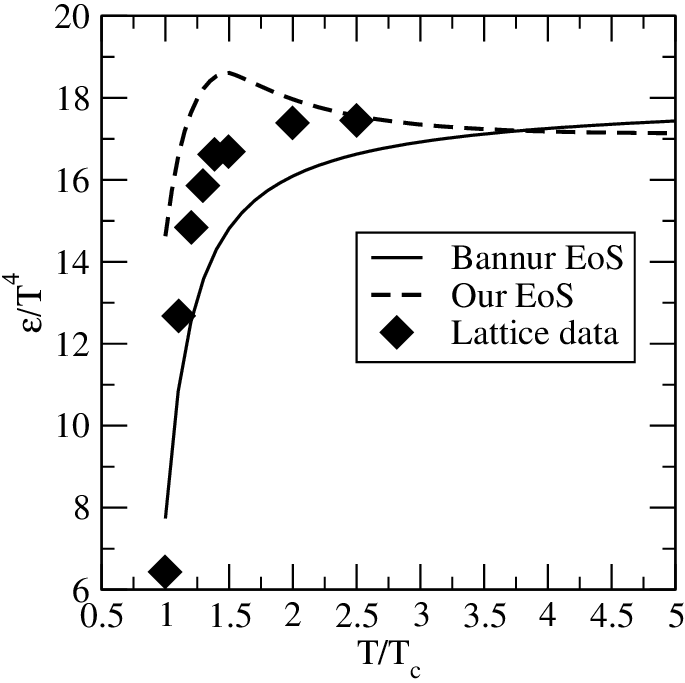}
\hspace{4mm}
\includegraphics[scale=.35]{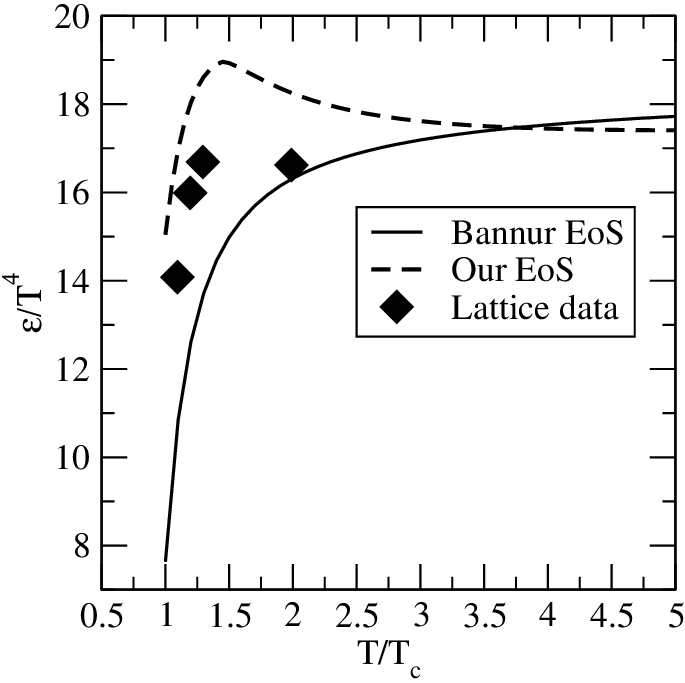}
\vspace{-1mm}
\caption{Variation of $\varepsilon/ T^4 $ as a function of $T/T_c$ for Bannur,
Our EoS (using quasi-particle Debye mass) and lattice results~\cite{banscqgp, boyd} for two
massless and one massive (2+1) extremely left, middle and extremely right for 4-flavour QGP for 
two different masses, $m/T$=0.4 and 0.2, respectively for the EoS2~\cite{kajantie}
where the notations are the same as in Fig.1.}
\vspace{26mm}
\end{figure*}
\section{Equation of States of different flavors in quasi-particle picture}
An extensive study of strong-coupled plasma in QED with proper modifications
to include colour degrees of freedom and the strong running coupling constant gives an expression for the energy density as a function of the plasma parameter can be
written as: 
\begin{equation} 
\varepsilon = \left( \frac{}{} 3 + u_{ex} (\Gamma) \right) 
\, n \, T \;, \end{equation}
Now, the scaled-energy density is written as in terms of ideal contribution
\begin{equation} 
e(\Gamma) \equiv \frac{\varepsilon}{\varepsilon_{SB}} 
= 1 + \frac{1}{3} u_{ex} (\Gamma) \quad,
\label{eps}
\end{equation}

 At sufficiently high temperature one must expect hadrons to “melt”, deconfining quarks and gluons. The exposure of new (color) degrees of freedom would then 
 be manifested by a rapid increase in entropy density, hence in pressure, with increasing temperature, and by a consequent change in the equation of state (EOS)~\cite{star}.
In this section we will find the pressure, energy density and speed of sound for pure gauge, 2-flavor , 3-flavor , (2+1)-flavor and 4-flavors QGP for EoS1 and EoS2.
To begin with first of all, we will calculate the energy density $\varepsilon (T)$ from Eq. (\ref{eps}) and using the thermodynamic relation, 
\begin{equation} 
\varepsilon = T \frac{dp}{dT} - P \quad, 
\end{equation}
we calculated the pressure as 
\begin{equation} \frac{P}{T^4} = \left( \frac{P_0}{T_0} + 3 a_f \int_{T_0}^T \, 
d\tau \tau^2 e(\Gamma(\tau)) \right) / T^3 \; , \label{eq:p} \end{equation} 
here $P_0$ is the pressure at some reference temperature $T_0$. This
temperature has been fixed with the values of pressure at critical 
temperature $275 MeV$, $175 MeV$, $155 MeV$ and $205 MeV$ for a particular 
system -pure gauge, 2-flavor, 3-flavor and 4-flavor QGP respectively. For the sake of comparison with the results of Bannur EoS we took the same value of
critical temperature as used in Bannur Model. Now, the speed of sound $c_s^2 (= \frac{dP}{d\varepsilon})$ can be calculated once we know the pressure $P$ and
energy density $\varepsilon$. 
In Fig. 1 and Fig. 2, we have plotted the variation of pressure ($P/T^4$) with
temperature ($T/T_c$) using EoS1 and EoS2 for pure gauge, 2-flavor and 3-flavor QGP along with Bannur EoS~\cite{banscqgp} and compared it with 
lattice results~\cite{banscqgp, boyd}.
For each flavor, $g_c$ and $\Lambda_T$ are adjusted to get a good fit 
to lattice results in Bannur Model. However, in our calculation we have fixed $P_0$ from the lattice data at the critical temperature $T_c$ for
each system as mentioned above, and there is no quantity to be fitted for predicting lattice results as done in Bannur case. 
Now, energy density $\varepsilon$, speed of sound $c_s^2$ etc. can be derived
since we had obtained the pressure, $P(T)$ .
In Fig. 3 and Fig. 4, we had plotted the energy density 
($\varepsilon / T^4$) with temperature ($T/ T_c$) using EoS1~\cite{zhai,arnold} and EoS2 for pure gauge, 2-flavor and 3-flavor QGP along with Bannur EoS~\cite{banscqgp}
and compared it with lattice result~\cite{banscqgp, boyd}. We observe that reasonably good fit is obtained without any extra parameters
for all three systems. As the flavor increases, the curves shifts to left.
In Fig. 5 and Fig. 6, the speed of sound, $c_s^2$ is plotted for all three systems, using EoS1 and EoS2 for pure gauge, 2-flavor and 3-flavor QGP along 
with Bannur EoS~\cite{banscqgp}. Since lattice results are available for only pure gauge, therefore comparison
has been checked for the above mentioned flavor only. Our flavored results matches excellent with the lattice results. We observe that 
 as the flavor increases $c_s^2$ becomes larger for both EoS1 and EoS2. All three curves shows similar behaviour, i.e, sharp rise near $T_c$ and then flatten to the
ideal value ($1/3$). However, in the vicinity of critical temperature, fits or predictions may not be good, especially for
energy density $\varepsilon$ and $c_s^2$ which strongly depends on variations of pressure $P$ with respect to temperature $T$. However, except for small 
region at $T=T_c$, our results are very good for all regions of $T > T_c$.
It is interesting to note that Peshier and Cassing~\cite{pe.3} also obtained similar results on
the dependence of plasma parameter $\Gamma$ in quasi-particle model 
and concluded that QGP behaves like a liquid, not weakly-interacting gas.
Now for the realistic case u and d quarks
have very small masses (5-10 MeV), strange quarks are having masses
150-200 MeV and charm quark with mass 1.5 GeV. 
Let $g_f$ counts the effective number of degrees of freedom of a 
massive Fermi gas. For a massless gas we have, of course, $g_f = n_f$.
  In Fig. 7-10, we have shown our results on (2+1)-flavors and 4-flavors QGP using EoS1 and EoS2 for pure gauge, 2-flavor and 3-flavor QGP and compared
it with Bannur EoS along with lattice data~\cite{ka.3,ka.4} and replotted the variation of $P(T)/T^4$ and energy density $\varepsilon (T)/T^4$ with temperature
$T/T_c$ for all systems.
This has been concluded that in the massless limit the deviations of pressure from the ideal gas value is larger in the presence of a heavier quark. 
This is in qualitative agreement with the observations.
We also calculate the thermodynamical quantities {\em viz.} pressure, screening energy density ($\epsilon_s$), the speed of sound etc. to study the hydrodynamical expansion 
of plasma and finally, to estimate the suppression of $J/\psi$ in nuclear collisions.
\begin{figure*}
\vspace{-1mm}
\includegraphics[scale=.40]{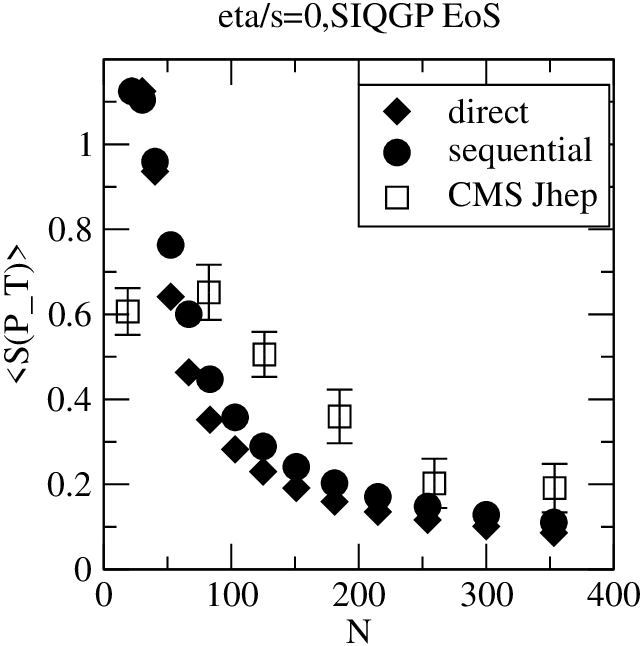}
\hspace{4mm}
\includegraphics[scale=.40]{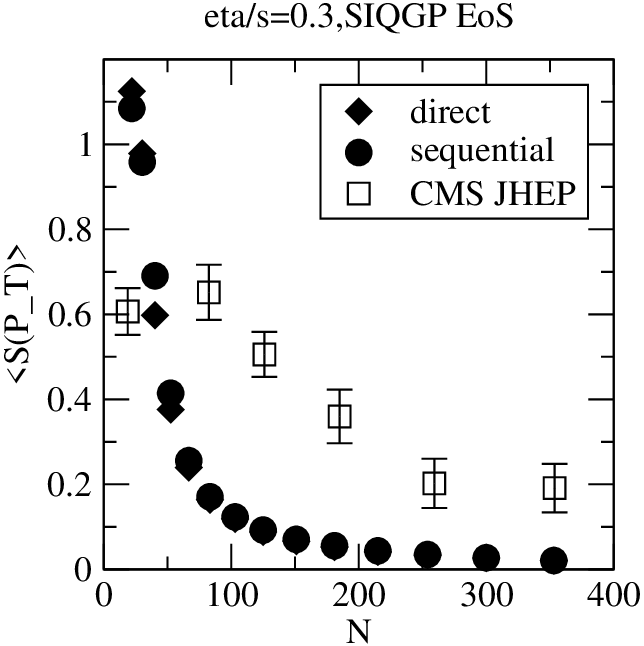}
\hspace{4mm}
\includegraphics[scale=.40]{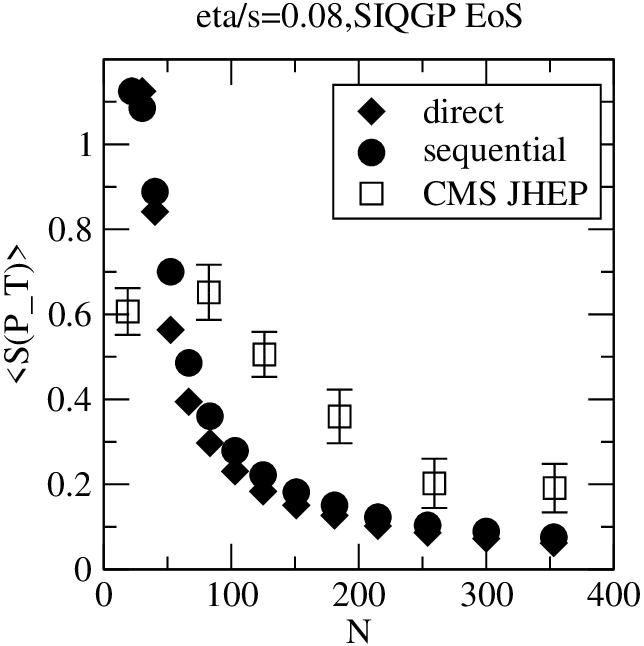}
\vspace{-1mm}
\caption{The variation of $\pt$ integrated survival probability
(in the range allowed by invariant $\pt$ spectrum of $J/\psi$ by the
CMS experiment) versus number of participants at mid-rapidity for the EoS1~\cite{kajantie}.
The experimental data (CMS JHEP)~\cite{expt1} are shown by the squares with error bars whereas
circles and diamonds represent with ($\sincl$) without ($\sdir$) sequential melting  using the values of $T_D$'s and related parameters
from Table I using SIQGP equation of state.}
\vspace{30mm}
\end{figure*}
\begin{figure*}
\vspace{2mm}
\includegraphics[scale=.40]{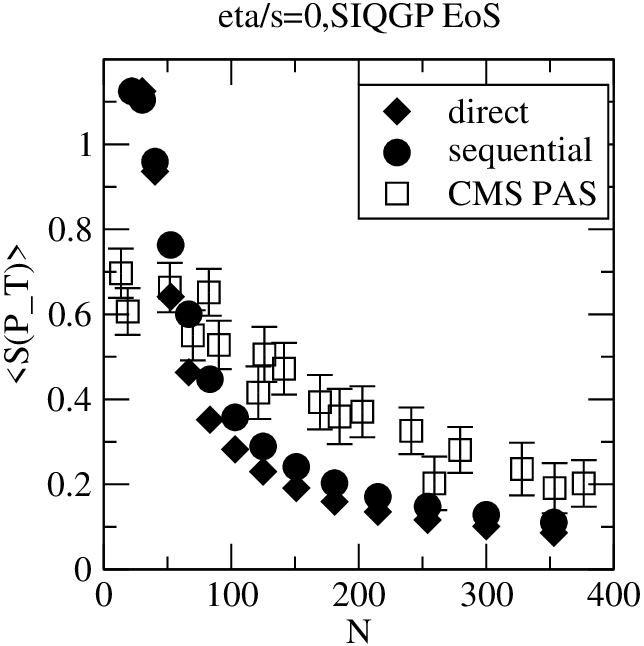}
\hspace{4mm}
\includegraphics[scale=.40]{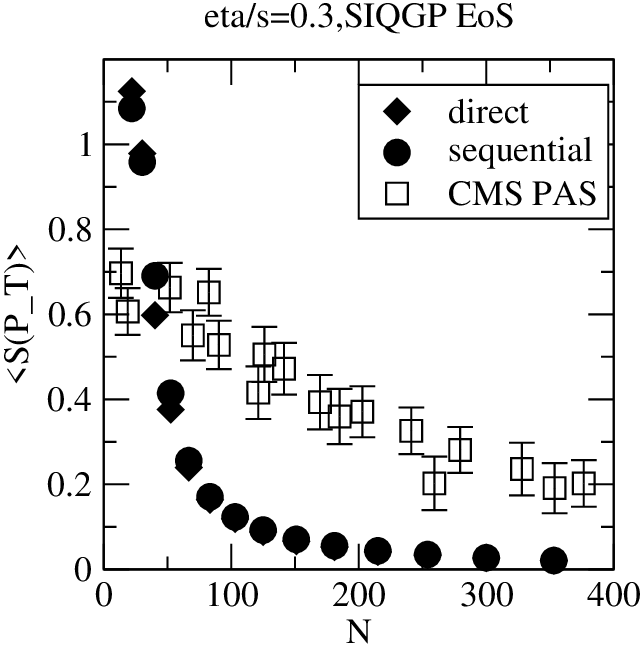}
\hspace{4mm}
\includegraphics[scale=.40]{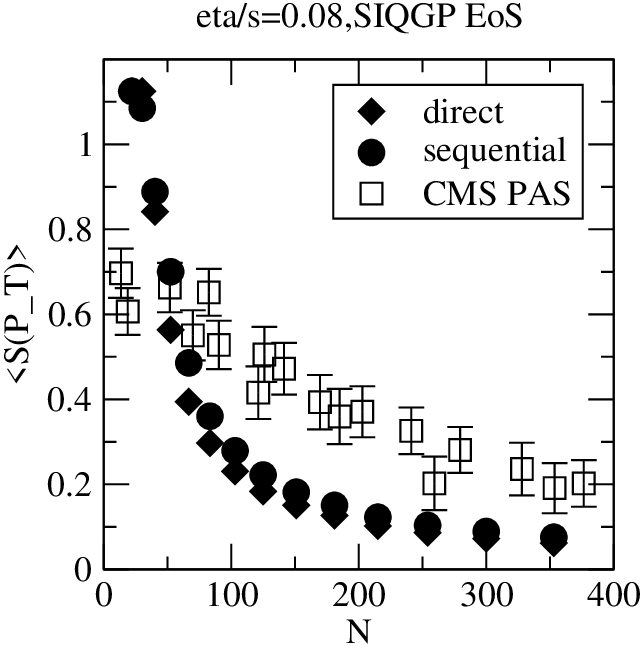}
\vspace{-1mm}
\caption{The variation of $\pt$ integrated survival probability
(in the range allowed by invariant $\pt$ spectrum of $J/\psi$ by the
CMS experiment) versus number of participants at mid-rapidity for the EoS1~\cite{zhai,arnold}.
The experimental data (CMS PAS)~\cite{expt2} are shown by the squares with error bars whereas
circles and diamonds represent with ($\sincl$) without ($\sdir$) sequential melting  using the values of $T_D$'s and related parameters from
Table I using SIQGP equation of state.}
\vspace{32mm}
\end{figure*}
\begin{figure*}
\vspace{-1mm}
\includegraphics[scale=.40]{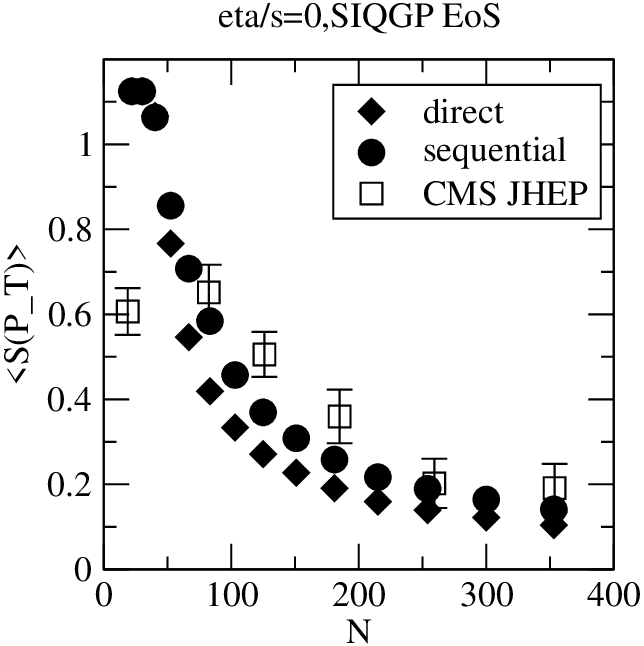}
\hspace{4mm}
\includegraphics[scale=.40]{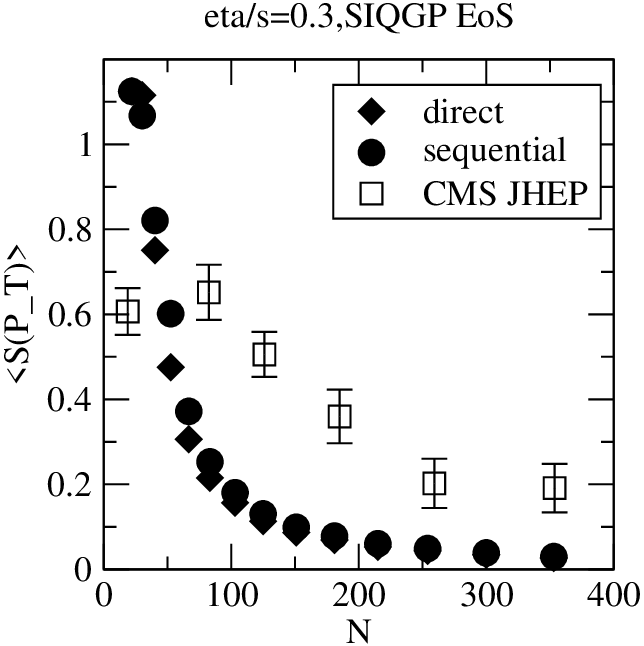}
\hspace{4mm}
\includegraphics[scale=.40]{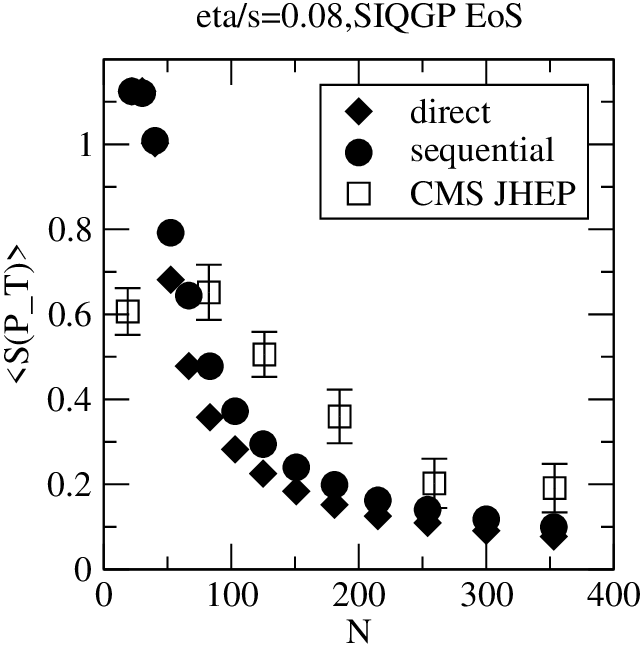}
\vspace{-1mm}
\caption{The variation of $\pt$ integrated survival probability
(in the range allowed by invariant $\pt$ spectrum of $J/\psi$ by the
CMS experiment) versus number of participants at mid-rapidity for the EoS2~\cite{kajantie}.
The experimental data (CMS JHEP)~\cite{expt1} are shown by the squares with error bars whereas
circles and diamonds represent with ($\sincl$) without ($\sdir$) sequential melting  using the values of $T_D$'s and related parameters
from Table II using SIQGP equation of state.}
\vspace{30mm}
\end{figure*}
\begin{figure*}
\vspace{2mm}
\includegraphics[scale=.40]{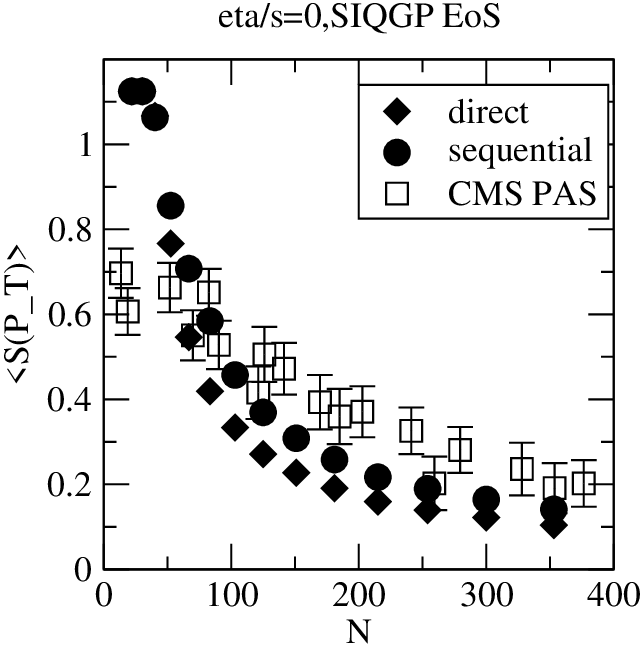}
\hspace{4mm}
\includegraphics[scale=.40]{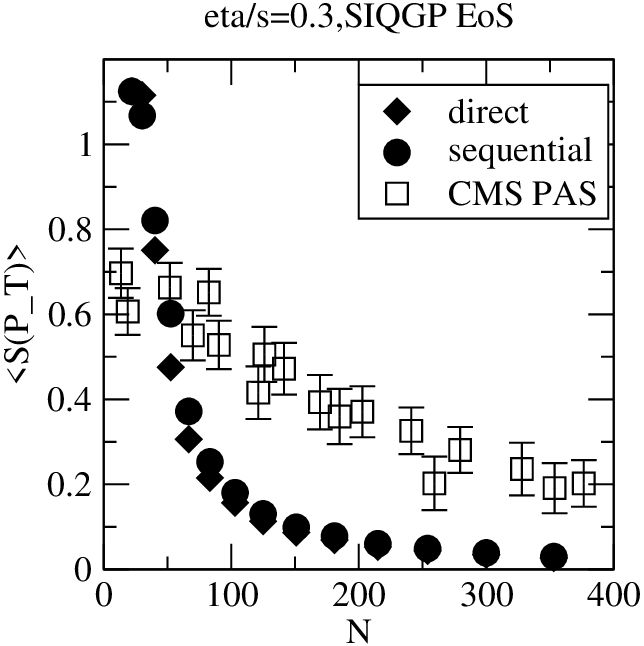}
\hspace{4mm}
\includegraphics[scale=.40]{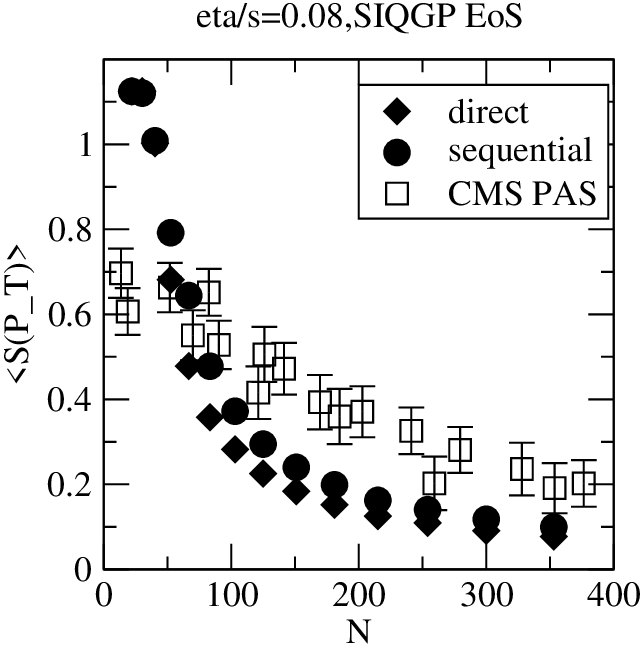}
\vspace{-1mm}
\caption{The variation of $\pt$ integrated survival probability
(in the range allowed by invariant $\pt$ spectrum of $J/\psi$ by the CMS experiment) versus number of participants at mid-rapidity for the EoS2\cite{zhai,arnold}.
The experimental data (CMS PAS)~\cite{expt2} are shown by the squares with error bars whereas
circles and diamonds represent with ($\sincl$) without ($\sdir$) sequential melting  using the values of $T_D$'s and related parameters from
Table II using SIQGP equation of state.}
\vspace{32mm}
\end{figure*}
\section{Survival probability of $c \bar c$ states}
To obtain the charmonium survival probability for an expanding QGP/QCD medium in the presence of viscous forces, the solution of equation of motion gives the time $\tau_s$ , which is estimated when the energy density drops to the screening energy density $\epsilon_s$ as
\begin{eqnarray}
\label{taus}
\tau_s(r)=\tau_i {\bigg[ \frac{\epsilon_i(r)-
\frac{4a}{3{\tilde{\tau}}_i^2}}{\epsilon_s-\frac{4a}{3{\tilde{\tau}}_s^2}}
\bigg]}^{1/1+c_s^2}
\end{eqnarray}
where $\epsilon_i(r)=\epsilon(\tau_i,r)$ and 
${\tilde{\tau}}_s^2$ is $(1-c_s^2)\tau_s^2$.
The critical radius $r_s$, is seen to mark the boundary of the region 
where the quarkonium formation is suppressed, can be obtained by
equating the duration of screening
$\tau_s(r)$ to the formation time $t_F=\gamma \tau_F$ for the quarkonium
in the plasma frame and is given by:
\begin{eqnarray}
\label{rs}
r_s= R_T { \left( 1- A \right)}^{\frac{1}{2}} \theta \left( 1-A \right)~,
\end{eqnarray}
The quark-pair will escape the screening region (and form quarkonium) 
if its position $\mathbf{r}$ and 
transverse momentum $\mathbf{p}_T$ are such that
\begin{equation}
\left| \mathbf{r}+\tau_F \mathbf{p}_T/M\right| \geq r_s.
\end{equation}
Thus, if $\phi$ is the angle between the vectors $\mathbf{r}$ and 
$\mathbf{p}_T$,
 then 
\begin{equation}
\cos \phi\,\geq\,\left[(r_s^2-r^2)\,M-\tau_F^2\,p_T^2/M\right]/
\left[2\,r\,\tau_F\,p_T\right],
\label{phi}
\end{equation}    
Here we choose $\alpha=0.5$ in our calculation as used in Ref.~\cite{chu}. Therefore the survival probability for the charmonium in QGP medium can be expressed 
as~\cite{mmish,chu} :
\begin{equation}
S(p_T,N_{part})=\frac{2(\alpha+1)}{\pi R_T^2}\int_0^{R_T}dr r \phi_{max}(r)\left\{1-\frac{r^2}{R_T^2}\right\}^{\alpha},
\end{equation}
where $\phi_{max}$ is the maximum positive angle~\cite{suppr}. In nuclear collisions, the $\pt$-integrated inclusive
survival probability of $J/\psi$ in the QGP/QCD medium becomes~\cite{satz,dpal}.
\begin{equation}
\langle S^{{}^{\rm incl}} \rangle = 0.6 {\sdir}_{J/\psi}
+0.3 {\sdir}_{{}_{\chic}}
+0.1 {\sdir}_{{}_{\psi^\prime}}
\end{equation}
\section{Results and Discussion}
Now we will discuss the physical understanding
of charmonium suppression due to screening in the deconfined medium
produced in relativistic nucleus-nucleus collisions. This involves a competition
of various time-scales involved in an expanding plasma. 
From the table I and II we observe that the value of 
$\epsilon_s$ is different for different charmonium 
states and varies from one EoS to other. 
If $\eps \gtrsim \epsilon_i$, then 
there will be no suppression at all i.e., 
survival probability, $\spt$ is equal to 1.
With this physical understanding we analyze our results,$\sinpt$ as a function of the number of participants $N_{{Part}}$ in an expanding QGP. 
At RHIC energy, $J/\psi$ yields have been resulted 
from a balance between annihilation of $J/\psi$'s due to hard, thermal 
gluons~\cite{xusat,gluon1} along with colour screening~\cite{mish,chu} 
and enhancement due to coalescence of uncorrelated $c\bar c$ 
pairs~\cite{grand,andro,thews} which are produced thermally at 
deconfined medium.
A detailed investigation of the scaling properties of $J/\psi$ suppression as a function 
of several centrality variables
would give valuable insights into the origin of the
observed effect~\cite{NA60}.
However, recent CMS data do not show a fully confirmed indication 
of $J/\psi$ enhancement except for the fact that 
$\langle p_T^2\rangle$ of the data and shape of rapidity-dependent 
nuclear modification factor 
$R_{AA}(y)$~\cite{expt1,expt2,adare,martin} show some characteristics of coalescence production.

In our analysis, we have employed the quasi-particle debye mass to determine the
dissociation temperatures for the charmonium states ($J/\psi$, $\psi^\prime$, $\chic$ 
etc.) in isotropic medium computed in table I and II for EoS1 and EoS2 respectively. On that dissociation temperature we
had calculated the screening energy densities, $\eps$ and the speed of sound $c_s^2$ which are also listed in the table I and II for both EoS1 and EoS2. 
These values will be used as inputs, to calculate $\sinpt$.
\begin{table}
\label{table1}
\centering
\caption{Dissociation temperature$T_D$ (for a 3-flavor QGP),
using quasi-particle debye mass for charmonium states, for EoS1.}
\vspace{3mm}
\begin{tabular}{|l|l|l|l|l|l|l|l|}
\hline
State &$\tau_F$  &$T_D$  &$c_s^2$(SIQGP)  &$c_s^2$(Id)  &$\epsilon_s$(SIQGP)  &$\epsilon_s$(Id) \\
\hline\hline
$J/\psi$ &0.89&  1.60 & 0.330 &1/3 & 9.94 &9.84 \\
\hline
$\psi'$ & 1.50& 1.29 & 0.302 &1/3 & 4.10 &4.09\\
\hline
$\chi_c$ &2.00& 1.40 & 0.320 &1/3 & 5.63 &5.61\\
\hline
\end{tabular}
\end{table}

\begin{table}
\label{table2}
\centering
\caption{Dissociation temperature$T_D$ (for a 3-flavor QGP),
using quasi-particle debye mass for charmonium states, for EoS2}
\vspace{3mm}
\begin{tabular}{|l|l|l|l|l|l|l|l|}
\hline
State &$\tau_F$  &$T_D$  &$c_s^2$(SIQGP)  &$c_s^2$(Id)  &$\epsilon_s$(SIQGP)  &$\epsilon_s$(Id) \\
\hline\hline
$J/\psi$ &0.89& 1.64 & 0.331 &1/3 & 11.05 &10.93\\
\hline
$\psi'$ & 1.50& 1.36 & 0.316 &1/3 & 4.99 &4.98\\
\hline
$\chi_c$ &2.00& 1.46 & 0.322 &1/3 & 6.75 &6.72\\
\hline
\end{tabular}
\end{table}

We have shown the variation of $\pt$-integrated survival probability
in the range allowed by invariant $\pt$ spectrum of $J/\psi$ in
CMS experiment with $N_{{Part}}$ at mid-rapidity and compared with the experimental data (CMS JHEP)~\cite{expt1} in Fig.11 and Fig.13 and (CMS PAS)~\cite{expt2} in
Fig.12 and Fig.14. For this we had used the values of $T_D$'s and related parameters from Table I and II using SIQGP equation of state for both EoS1 and EoS2 .

We find that the survival probability of
sequentially produced $J/\Psi$ is slightly higher compared to the directly produced $J/\Psi$ and is closer to the experimental results.  
The smaller value of screening energy density $\epsilon_s$
causes an increase in the screening time and results in 
more suppression to match with the CMS results at LHC.
 We have also plotted the pressure, energy density and
speed of sound for pure gauge, 2-flavor, 3-flavor,(2+1)-flavors and 4-flavors QGP for both EoS1 and EoS2 in fig.1-10 where we have employed QP
EoS (QP EoS is the equation of state calculated by using quasi-particle debye mass) along with the Bannur EoS. Here we observe that the results 
of various equation of states coming by incorporating the quasi-particle Debye mass increases sharply.

\section{Conclusion}
We studied the equation of state for strongly 
interacting quark-gluon plasma in the framework of strongly coupled plasma 
with appropriate modifications to take account of color and flavor degrees
of freedom and QCD running coupling constant.
In addition, we incorporate the nonperturbative effects in terms 
of nonzero string tension in the 
deconfined phase, unlike the Coulomb interactions alone in the
deconfined phase beyond the critical temperature.
Our results on thermodynamic observables
{\em viz.} pressure, energy density, speed of sound etc. nicely fit the
results of lattice equation of state with gluon, massless and as
well {\em massive} flavored plasma. In Fig.1-10 we see that the results coming out by using quasi-particle Debye mass increases sharply as the temperature increases.
Now by using quasi-particle Debye mass we estimated the centrality dependence of charmonium suppression in an expanding dissipative
strongly interacting
QGP produced in relativistic heavy-ion collisions as shown in Fig.11-14 for both EoS1 and Eos2. We find that the survival probability of
sequentially produced $J/\Psi$ is slightly higher compared to the directly produced $J/\Psi$ and is closer to the experimental results. 
The smaller value of screening energy density $\epsilon_s$ causes an increase in the screening time and results in more suppression to 
match with the experimental results.

At LHC energies, the inclusive $J/\psi$ yield contains a significant non-prompt contribution from b-hadron decays~\cite{lhcb,cms}. 
For the lower value of $\eta/s$ we observe that our predictions are closer to the experimental ones.
\section*{Acknowledgement}
VKA acknowledge the Science and Engineering research Board (SERB) Project No. {\bf EEQ/2018/000181} New Delhi for financial support. We record our sincere gratitude to the people of India for their generous support for the research in basic sciences.

\end{document}